\pdfoutput=1
\documentclass[twocolumn,nobibnotes,nofootinbib,raggedbottom,10pt]{revtex4-2}
\usepackage{hyperref,graphicx,array,multirow,color,amsfonts,amsmath,mathrsfs,titlesec,shuffle,tikz,ifthen}
\widowpenalty=500\clubpenalty=1000\unitlength=1mm\textheight=8.95in\hypersetup{pdftitle={},pdfcreator={},linkcolor=[rgb]{0.15,0.35,0.75},colorlinks=true,citecolor=[rgb]{0.675,0,0.2},urlcolor=[rgb]{0.15,0.35,0.65}}
\renewcommand{\footnoterule}{\kern -0.75ex\noindent\rule{\linewidth}{0.5pt}\\\vspace{5.5ex}}

\newcommand{\eq}[1]{\vspace{-0.5pt}\begin{equation}#1\vspace{-0.5pt}\end{equation}}
\newcommand{\fwbox}[2]{\text{\makebox[#1][c]{$\hspace{-150pt}\displaystyle#2\hspace{-150pt}$}}}
\newcommand{\fwboxL}[2]{\text{\makebox[#1][l]{$#2$}}}
\newcommand{\fwboxR}[2]{\text{\makebox[#1][r]{$#2$}}}
\newcommand{\equivR}{\fwbox{15.5pt}{\hspace{-0pt}\fwboxR{0pt}{\raisebox{0.47pt}{\hspace{1.25pt}:\hspace{-4pt}}}=\fwboxL{0pt}{}}}
\newcommand{\equivL}{\fwbox{15.5pt}{\fwboxR{0pt}{}=\fwboxL{0pt}{\raisebox{0.47pt}{\hspace{-4pt}:\hspace{1.25pt}}}}}

\newcommand{\bigger}[1]{\raisebox{-0.95pt}{\scalebox{1.25}{$#1$}}}
\newcommand{\mi}{\raisebox{0.75pt}{\scalebox{0.75}{$\hspace{-0.5pt}\,-\,\hspace{-0.5pt}$}}}
\renewcommand{\pl}{\raisebox{0.75pt}{\scalebox{0.75}{$\hspace{-0.5pt}\,+\,\hspace{-0.5pt}$}}}
\renewcommand{\phi}{\varphi}
\renewcommand{\bar}{\overline}
\renewcommand{\hat}{\widehat}
\renewcommand{\tilde}{\widetilde}
\newcommand{\ab}[1]{\langle #1\rangle}
\renewcommand{\sb}[1]{[ #1]}

\newcommand{\dzero}{{\color{dim}0}}

\DeclareMathSymbol{\shortminus}{\mathbin}{AMSa}{"39}
\renewcommand{\mi}{\fwbox{1ex}{\shortminus}}
\renewcommand{\pl}{\fwbox{1ex}{\fwbox{0pt}{\shortminus}\fwbox{0pt}{\rotatebox[origin=c]{90}{$\shortminus$}}}}

\renewcommand{\r}[1]{{\color{hred}#1}}
\newcommand{\g}[1]{{\color{hgreen}#1}}
\renewcommand{\b}[1]{{\color{hblue}#1}}
\renewcommand{\t}[1]{{\color{hteal}#1}}

\definecolor{hblue}{rgb}{0.2353,0.2353,0.60}
\definecolor{hred}{rgb}{0.575,0.0,0.225}
\definecolor{hgreen}{rgb}{0.0,0.5,0.4}
\definecolor{hteal}{rgb}{0.0,0.545,0.7451}
\definecolor{hpurple}{rgb}{0.4431,0.2862,0.7960}

\definecolor{dualblue}{rgb}{0.169,0.22,0.565}
\definecolor{hblue}{rgb}{0,0,0.575}
\definecolor{hred}{rgb}{0.575,0.0,0.225}
\definecolor{dim}{rgb}{0.55,0.55,0.55}
\definecolor{deemph}{rgb}{0.25,0.25,0.25}
\definecolor{rindou1}{rgb}{0.4431,0.2862,0.7960}
\definecolor{rindou2}{rgb}{0.0078,0.1215,0.4392}
\definecolor{lapis}{rgb}{0.0.0470,0.2941,0.5568}
\definecolor{emerald}{rgb}{0.31, 0.78, 0.47}
\definecolor{pinegreen}{rgb}{0.0, 0.47, 0.44}
\definecolor{jade}{rgb}{0.0, 0.66, 0.42}
\definecolor{teal}{rgb}{0.0, 0.5, 0.5}

\begin{document}
\title{\texorpdfstring{The Supersymmetry of Cuts in Pure Gauge Theory and Gravity\\[-22pt]~}{The Supersymmetry of Cuts in Pure Gauge Theory and Gravity}}
\author{Jacob~L.~Bourjaily}
\affiliation{Institute for Gravitation and the Cosmos, Department of Physics,\\Pennsylvania State University, University Park, PA 16802, USA}
\begin{abstract} 
At tree-level, scattering amplitudes involving only gluons or gravitons are unaffected by supersymmetry, allowing them to be efficiently encoded by and extracted from those of maximally supersymmetric ($\mathcal{N}\!=\!4,8$) theories. This fails beyond tree-level, of course, but much less than would be expected. We show that all the leading singularities of (sub-maximally or) non-supersymmetric theories can be organized into \emph{generalized} superfunctions, in terms of which all helicity components can be effectively encoded. These functions differ from those of maximally supersymmetric theories by an extent determined by loop-order---broken into a sum over $2^L$ supersymmetric pieces.
\\[-12pt]
\vspace{-10pt}
\end{abstract}
\maketitle

\vspace{-16pt}\section{Introduction}\label{introduction_section}\vspace{-14pt}
%
The use of supersymmetry has long proven extremely useful for the computation of amplitudes in (sub-maximally or) \emph{non}-supersymmetric theories. At tree-level, maximal supersymmetry can be viewed entirely as a book-keeping device, as gluons or gravitons cannot interact with other states except through loops. This has been put to powerful use in a variety of public code and computational tools for trees (see \emph{e.g}~\cite{Dixon:2010ik,Bourjaily:2010wh,Bourjaily:2023uln}). Beyond tree-level, however, amplitudes in theories with different degrees of supersymmetry obviously differ. 

Perturbative scattering amplitudes in gauge theory can be computed in terms of their leading singularities according to generalized unitarity \mbox{\cite{Bern:1993tz,Bern:1993mq,Bern:1994zx,Bern:1994cg,Bern:1995db,Bern:1995ix,Bern:1996ja}} (see also~\mbox{\cite{Roiban:2004ix,Berger:2006ci,Berger:2006vq,Ellis:2006ss,Ellis:2007br,Berger:2008sj,Ellis:2008ir,Giele:2008ve,Degrande:2018neu,Bourjaily:2020qca}}), which has led to impressive improvements in both our understanding of scattering amplitudes and our ability to compute them (see~\emph{e.g.}~\mbox{\cite{Su:2006vs,Xiao:2006vt,Dixon:2009uk,Abreu:2018aqd,Abreu:2019rpt}}). This is no more apparent than in the case of amplitudes in theories with maximal supersymmetry (see~\emph{e.g.}~\mbox{\cite{Bourjaily:2013mma,Bourjaily:2015jna,Bourjaily:2017wjl,Abreu:2018aqd,Bourjaily:2018omh,Bourjaily:2019gqu,Abreu:2019rpt,Bourjaily:2021iyq}}), integrands of which led to the discovery of on-shell recursion relations at tree-level \mbox{\cite{BCF,BCFW,ArkaniHamed:2008yf,Elvang:2011fx}} and beyond \mbox{\cite{ArkaniHamed:2010kv,Bourjaily:2023apy}}. 

Despite this tremendous progress, there is considerably less known about amplitudes in pure (or even less supersymmetric) gauge theory and gravity. Partly, this can be attributed to the difficulties of considering each of the $\binom{n}{k}$ `helicity' (or super-)\emph{component} amplitudes separately (see~\emph{e.g.}~\cite{Bourjaily:2021ujs}).\\[-5pt]

In this Letter, we show that the full power of supersymmetric book-keeping can be preserved beyond tree-level. Specifically, we show that any $L$-loop leading singularity\footnote{Specifically, leading singularities constructed in terms of four-dimensional tree amplitudes---not those whose definition requires (\emph{e.g.} dimensional) regularization.} of pure (or less-supersymmetric) gauge or gravity theory can be expressed in terms of \emph{generalized} superfunctions which simultaneously encode all the particular configurations of external component states related by supersymmetry. In notation clarified below, an $L$-loop leading singularity of N${}^{k\mi2}$MHV-degree in a theory $\mathcal{N}$ supersymmetries can be expressed in terms of functions of the form\\[-12pt]
\vspace{-3pt}\eq{\mathfrak{f}_{\sigma}^{{\mathcal{N}}}(\vec{p},\tilde\eta,\theta)\equivR f_\sigma(\vec{p})\!\times\!\delta^{k\times\mathcal{N}}\!\big(C\!\cdot\!\tilde\eta\big)\hspace{-2pt}\sum_{\vec{s}\in[2]^L}\hspace{-4pt}\delta^{k\times\bar{\mathcal{N}}}\!\big(D_{\vec{s}}\!\cdot\!\theta\big)\vspace{-5pt}\label{archetype_less_superfunction}}
where $\sigma{=}1,2$ for sYM and sGR, respectively; $f_{\sigma}(\vec{p})$ is an ordinary function of on-shell momenta; $C$ and each of the $D_{\vec{s}}$ are $k\!\times\!n$ matrices of ordinary functions; for each $a\!\in\![n]$ and $I\!\in\![\mathcal{N}]$, $\tilde\eta_a^I$ is an auxiliary Grasmman parameter used to connect states related to the highest-helicity $h\!=\!{+}\sigma$ state by supersymmetry, and 
\vspace{-4pt}\eq{\theta_a\!\equivR\!\prod_{\bar{I}\in[\bar{\mathcal{N}}]}\tilde\eta_a^{\bar{I}},\vspace{-10pt}\label{thetas_defined}}
with $[\bar{\mathcal{N}}]\!\equivR\![4\sigma]\backslash[\mathcal{N}]$ introduced to label states connected to these by parity. Notice that in (\ref{archetype_less_superfunction}), the function $f_{\sigma}(\vec{p})$ does not depend on the degree of supersymmetry, but will be different for sYM and sGR; the matrices $C$ and $D_{\vec{s}}$ are the same for both theories.

Generalized superfunctions are defined in terms of an $L$-loop graph with vertices denoting tree-amplitudes with $L$ marked internal lines---the cutting of which would result in a tree of tree-amplitudes; for example, 
\vspace{-10pt}\eq{\begin{array}{@{}c@{}}\includegraphics[scale=1]{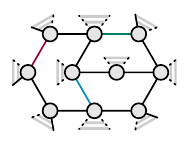}\end{array}\bigger{\Rightarrow}\begin{array}{@{}c@{}}\includegraphics[scale=1]{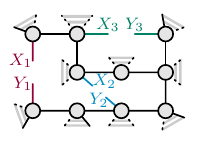}\end{array}\fwboxL{0pt}{\!\!.}\vspace{-10pt}}
The cut diagram will have $(n{+}2L)$ legs and N${}^{k\mi2\pl L}$MHV-degree; because it is built entirely from tree amplitudes for which supersymmetric bookkeeping can be used it can be defined independent of the degree (or existence) of supersymmetry in terms of ordinary (maximally-supersymmetric) superfunctions. From these, all component amplitudes can be extracted, including those involving only the subset of states relevant to $\mathcal{N}\!\geq\!0$ sYM or sGR to define the new functions. 

The distinction between theories with different degrees of supersymmetry, therefore, arises entirely through the restricted set of states to be summed for each pair of legs sewn together:
\vspace{-5pt}\eq{\sum_{\substack{\text{conjugate}\\\text{states }\r{X},\r{Y}}}\hspace{-5pt}\bigger{\Rightarrow}\int\!\!\big[d\theta_{\!\r{X}}{+}d\theta_{\hspace{-0.5pt}\r{Y}}\!\big]\!\!\!\!\prod_{I\in[\mathcal{N}]}\!\!\!\big(d\tilde\eta_{\r{X}}^{I}{+}d\tilde\eta_{\r{Y}}^{I}\big)\,.\vspace{-5pt}\label{general_state_sum}}
In this state sum, the Grassmann integral involving $d\theta_{\r{X}}$ includes all those states for which $|\r{X}\rangle$ is connected to $|\r{X}\rangle^{\mi\sigma}$ by supersymmetry, and similarly for the integral over $d\theta_{\r{Y}}$. Choosing one parity for each pair of legs to be merged results in one of the $2^L$ terms in (\ref{archetype_less_superfunction}):
\vspace{-5pt}\eq{f_\sigma(\vec{p})\!\times\!\delta^{k\times\mathcal{N}}\!\big(C\!\cdot\!\tilde\eta\big)\hspace{-0pt}\hspace{-0pt}\delta^{k\times\bar{\mathcal{N}}}\!\big(D_{\vec{s}}\!\cdot\!\theta\big).\vspace{-4pt}}
This particular contribution depends sensitively on the choice of edges cut in the original diagram; only when all $2^L$ terms are summed will the resulting generalized superfunction be independent of this choice. 

It is worth noting that because the cut diagrams are manifestly tree-graphs of tree-amplitudes, they necessarily admit a plane embedding. The ordering of legs on the boundary will depend on the orderings of partial amplitudes at each vertex---which can be defined unambiguously (if conventionally) for both sYM \cite{DelDuca:1999rs,Arkani-Hamed:2014bca} and sGR \cite{Paranjape:2023qsq,Koefler:2024pzv,Lemmon:2025dhq,Herrmann:2016qea,Heslop:2016plj,Bourjaily:2023ycy}. Thus, they are definable in terms of \emph{positroid} varieties \cite{ArkaniHamed:book,Bourjaily:2012gy,Bourjaily:2016mnp}. This is not terribly surprising: any non-planar on-shell diagram can be understood as constructed from a plane-embedded tree upon merging legs which need not be adjacent. But it is striking that the vocabulary of \emph{planar} diagrams suffices not only to describe the various \emph{non-planar} leading singularities of sYM, but also of pure YM and pure GR.

\vspace{-16pt}\section{Formalism: Superfunctions and Coherent States}\vspace{-14pt}
Let us briefly review the notation used to describe superfunctions of $\mathcal{N}$-supersymmetric Yang-Mills and gravity theories (`sYM' and `sGR'). Let $\tilde{Q}_I^{\dot{\alpha}}$ for $I\!\in\![\mathcal{N}]$ be generators of supersymmetry (if any exist). For each external particle indexed by $a\!\in\![n]$, with on-shell momentum $p_a^{\alpha\dot{\alpha}}\!\equivR p_a^\mu\sigma_\mu^{\alpha\dot{\alpha}}\equivL\lambda_a^\alpha\tilde\lambda_a^{\dot{\alpha}}$, there will be a variety of particular states related by supersymmetry/parity to those with the highest helicity $h{=}{+}\sigma$ with $\sigma{=}1,2$ for sYM and sGR, respectively. We gather these states into a \emph{coherent state} \cite{Arkani-Hamed:2008owk} as follows. 

For each particle we introduce $4\sigma$ auxiliary Grassmann parameters $\tilde\eta_a^I$ with $I\!\in\![4\sigma]{=}[\mathcal{N}]\bigger{\cup}\big([4\sigma]\backslash[\mathcal{N}]\big)$---regardless of the degree of supersymmetry---to be used for book-keeping. We define a spinor $\tilde{\lambda}^{\dot{\alpha}}_{\bar{a}}$ such that $\sb{\bar{a}\,a}\!\equivR\!\epsilon_{\dot{\alpha}\dot{\beta}}\tilde\lambda_{\bar{a}}^{\dot{\alpha}}\tilde\lambda_a^{\dot{\beta}}\!\equivR\!1$, and for each supersymmetry generator indexed by $I\!\in\![\mathcal{N}]$ we define $\tilde{Q}^a_{I}\!\equivR\!\epsilon_{\dot{\alpha}\dot{\beta}}\tilde{Q}_I^{\dot{\alpha}}\tilde\lambda^{\dot{\beta}}_{\bar{a}}$. The coherent state `$|a\rangle$' is then defined to be
\vspace{-4pt}\eq{|a\rangle\equivR\mathrm{exp}\big(\tilde{Q}_I^a\tilde\eta_a^I\big)\Big(|a\rangle^{\!\pl\sigma}{+}\theta_a\,|a\rangle^{\frac{\mathcal{N}}{2}\mi\sigma}\Big)\label{coherent_state_defined}\,\vspace{-4pt}}
with $\theta_a$ defined in (\ref{thetas_defined}) in terms of the subset of $\tilde\eta_a^{I}$'s with indices $I\!>\!\mathcal{N}$. From the coherent state, particular \emph{component} states can be extracted via Grassmann integration. For example, the state $|a\rangle^{\mi\sigma}$ would be obtained from (\ref{coherent_state_defined}) via\\[-10pt]
\vspace{-4pt}\eq{|a\rangle^{\mi\sigma}=\int\!\!d\tilde\eta_a^{1}\cdots d\tilde\eta_a^{4\sigma}\,\,|a\rangle=\int\!\!\Big(\!d\tilde\eta_a^1\cdots d\tilde\eta_a^{\mathcal{N}}\!\Big)d\theta_a\,\,|a\rangle\,.\vspace{-4pt}\label{projection_of_minus}}

Superfunctions are defined directly in terms of coherent states, and therefore depend not just on the momenta of the particles but also $\tilde\eta$'s and $\theta$'s used to package the various states in the theory together. Terms that arise from on-shell recursion of tree amplitudes, say, take the form 
\vspace{-8pt}\eq{\mathfrak{f}^{\mathcal{N}}_\sigma(\vec{p},\tilde\eta,\theta)=f_\sigma(\vec{p})\times\delta^{k\times\mathcal{N}}\!\big(C\!\cdot\!\tilde\eta\big)\delta^{k\times\bar{\mathcal{N}}}\!\big(C\!\cdot\!\theta\big)\vspace{-4pt}}
where the Grassmann $\delta$-functions depend on a single $(k\!\times\!n)$ matrix $C\!\equivR\!\{c^\alpha_a\}_{a\in[n],\alpha\in[k]}$; they are simply polynomials (following from the definition of Grassmann integration) defined by
\vspace{-4pt}\eq{\begin{split}\delta^{k\times\mathcal{N}}\!\big(C\!\cdot\!\tilde\eta\big)\!&\equivR\!\delta^{k\times\mathcal{N}}\!\Big(\hspace{-2pt}\sum_{a\in[n]}\hspace{-2pt}c_a^{\alpha}\tilde\eta_a^{I}\Big){=}\!\!\prod_{\substack{\alpha\in[k]\\I\in[\mathcal{N}]}}\!\Big(\hspace{-2pt}\sum_{a\in[n]}\hspace{-2pt}c_a^{\alpha}\tilde\eta_a^{I}\Big)\\[-12pt]
&\hspace{-28.pt}=\!\!\prod_{I\in[\mathcal{N}]}\!\!\left[\sum_{\hspace{-0pt}a_1<\cdots<a_k\hspace{-0pt}}\hspace{-10pt}\big(\tilde\eta_{a_1}^{I}\cdots\tilde\eta_{a_k}^I\big)\mathrm{det}(c_{a_1},\ldots,c_{a_k})\right]\!,\\[-12pt]\end{split}}
and
\vspace{-4pt}\eq{\begin{split}\delta^{k\times\bar{\mathcal{N}}}\!\big(C\!\cdot\!\theta\big)\!&\equivR\!\!\prod_{\substack{\alpha\in[k]}}\!\Big(\sum_{a\in[n]}c_a^{\alpha}\theta_a\Big)\\[-4pt]
&\hspace{-30pt}\fwboxL{0pt}{=\sum_{\smash{\hspace{-3.5pt}a_1<\cdots<a_k\hspace{-3.5pt}}}\hspace{-10pt}\big(\theta_{a_1}\!\!\cdots\theta_{a_k}\big)\mathrm{det}(c_{a_1},\ldots,c_{a_k})^{4\sigma\mi\mathcal{N}}}\\[-4pt]
&\hspace{-48pt}\fwboxL{210pt}{\hspace{19pt}\equivR\!\!\sum_{\hspace{-0pt}a_1<\cdots<a_k\hspace{-0pt}}\hspace{2pt}\!\!\prod_{\smash{\bar{I}\in[\bar{\mathcal{N}}]}}\hspace{-5.5pt}\Big(\!\big(\tilde\eta_{a_1}^{\bar{I}}\!\!\cdots\tilde\eta_{a_k}^{\bar{I}}\big)\mathrm{det}(c_{a_1},\ldots,c_{a_k})\!\Big).}\\[-5pt]
\end{split}}

Suppose one wanted to extract from the superfunction the \emph{component function}, denoted $f^{\vec{a}}_{\sigma}$, involving $({-}\sigma)$-helicity states labelled by $\{a_1,\ldots,a_k\}$, with all others having $({+}\sigma)$-helicity. This could be extracted analogously to what was done in (\ref{projection_of_minus}):
\vspace{-8pt}\eq{\begin{split}f^{\vec{a}}_{\sigma}&=\int\!\!\!\prod_{\alpha\in[k]}\!\!\Big(d\tilde\eta_{a_\alpha}^1\!\!\cdots d\tilde\eta_{a_\alpha}^{4\sigma}\Big)\mathfrak{f}^{\mathcal{N}}_{\sigma}(\vec{p},\tilde\eta,\theta)\\
&=f_\sigma(\vec{p})\,\,\mathrm{det}(c_{a_1},\ldots,c_{a_k})^{4\sigma}\,.
\end{split}\vspace{-12pt}}
For $\mathcal{N}\!<\!4\sigma$ supersymmetries, each superfunction---which is captured by a mere $\mathcal{O}(n\!\times\!k)$ particular functions---simultaneously encodes $\binom{n}{k}^{\mathcal{N}\pl1}$ distinct component functions. This concision can be understood from the manifestation of supersymmetric Ward identities \cite{Elvang:2009wd}. This is why it is computationally advantageous to use a maximally-supersymmetric formalism (when possible) to encode even non-supersymmetric amplitudes: the number of helicity amplitudes, $\mathcal{O}(n^{k})$, is much larger than the $\mathcal{O}(n\!\times\!k)$ functions required to specify $\mathfrak{f}$.

\vspace{-16pt}\section{Leading Singularities of YM/GR at One Loop}\vspace{-14pt}
\subsection{(Super-)\emph{Component}-Level Formulae}\vspace{-14pt}
Consider a one loop leading singularity relevant to MHV amplitudes in pure YM/GR:
\vspace{-12pt}\eq{\begin{array}{@{}c@{}}\includegraphics[scale=1]{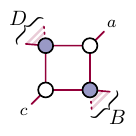}\end{array}\label{eg_mhv_one_loop}\vspace{-12pt}}
The value of this leading singularity---given by the product of the amplitudes at each vertex, integrated over the on-shell phase space of internal particles\footnote{This integral is trivial for all (polylogarithmic) leading singularities: it is entirely localized by momentum conservation at each vertex amplitude.}, and summed over possible gluon/graviton helicities of internal states \cite{ArkaniHamed:book}---clearly depends on the set of internal states being summed. Even for maximal supersymmetry, the states that may appear in the loop depends on the external states chosen (dictated by $R$-symmetry); for many cases involving external gluons/gravitons, only gluons/gravitons can appear inside the loop. This happens, for example, for helicity components
\vspace{-12pt}\eq{\raisebox{-27pt}{\includegraphics[scale=1]{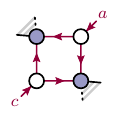}},\raisebox{-27pt}{\includegraphics[scale=1]{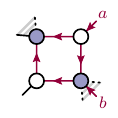}},\raisebox{-29.35pt}{\includegraphics[scale=1]{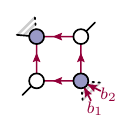}}\fwboxL{0pt}{\!.}\label{eg_mhv_one_loop_safe_components}\vspace{-12pt}}
For each of these components, the leading singularity (\ref{eg_mhv_one_loop}) will therefore be identical to those of the maximal supersymmetric theory. (In the figures above, all unlabelled legs have outgoing helicity flow.) However, the if the external helicities are chosen so that an oriented loop of internal helicity can flow, then all the component states that exist in the theory will contribute, resulting in different expressions depending on the degree of supersymmetry. For this leading singularity topology, this will happen for components of the form
\begin{align}\nonumber\\[-28pt]\hspace{-5pt}\raisebox{-26pt}{\includegraphics[scale=1]{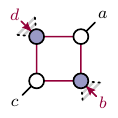}}
&\equivR
\raisebox{-26pt}{\includegraphics[scale=1]{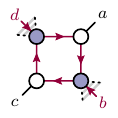}}{+}\raisebox{-26pt}{\includegraphics[scale=1]{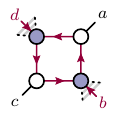}}\label{eg_mhv_one_loop_differing_components}\\[-5pt]
&=f_{\sigma}(\vec{p})\ab{b\,d}^{4\sigma}\hspace{-2pt}\big(\hspace{-1pt}\phi(a,\r{b};c,\r{d})^{4\sigma}\hspace{-2pt}{+}\phi(\r{d},a;\r{b},c)^{4\sigma}\hspace{-1pt}\big)\nonumber\\[-20pt]\nonumber
\end{align}
where $f_1(\vec{p}){=}\delta^{2\times2}\!\big(\lambda\hspace{-1pt}\!\cdot\hspace{-1pt}\!\tilde\lambda\big)/(\ab{1\,2}\!\cdots\!\ab{n\,1})$ for Yang-Mills\footnote{This term may be dressed color tensors and re-ordered according to partial amplitudes appearing at the corners. The function $f_2(\vec{p})$ for gravity can easily be constructed from the amplitudes involved (\emph{e.g.}~using Hodges' formula~\cite{Hodges:2012ym}), but an analytic expression would be too cumbersome to give here.} and the cross ratio $\phi$ is defined by
\vspace{-4pt}\eq{\phi(a,b;c,d)\equivR\frac{\ab{a\,b}\ab{c\,d}}{\ab{a\,c}\ab{b\,d}}\,.\vspace{-4pt}}
Notice that $\phi(\r{d},a;\r{b},c){=}1{-}\phi(a,\r{b};c,\r{d})$. 

It is not hard to upgrade the particular component formula (\ref{eg_mhv_one_loop_differing_components}) for external gluons/gravitons to a \emph{super}component formula valid for any degree of supersymmetry: allowing the legs $\{\r{b},\r{d}\}$ to be merely those connected by $\mathcal{N}$-supersymmetry to (${-}\sigma$)-helicity gluons/gravitons, we would have\\[-12pt]
\vspace{-2pt}\eq{\hspace{-115pt}f_{\sigma}(\vec{p})\,\delta^{2\times\mathcal{N}}\!\big(\lambda\!\cdot\!\tilde\eta\big)\big(\phi(a,\r{b};c,\r{d})^{4\sigma\mi\mathcal{N}}\hspace{-4pt}{+}\phi(\r{d},a;\r{b},c)^{4\sigma\mi\mathcal{N}}\big).\hspace{-100pt}\label{general_n_2me_ls_olde}\vspace{-2pt}}

This expression is quite similar to what one would find using the Grassmannian formalism of \cite{ArkaniHamed:book}, for which a perfectly oriented, trivalent on-shell diagram corresponds to a function differing from that of maximal supersymmetry by a factor determined by face variables. 

But this description of on-shell functions still requires a \emph{distinct} expression for all $\binom{n}{k}$ of its \mbox{(super-)}components. Enumerating all such choices for all possible leading singularities even at one loop would require significant organizational efforts. Generalized superfunctions completely eliminate the need for such enumerations. 

As we will describe momentarily, the description of the leading singularity (\ref{eg_mhv_one_loop}) in terms of generalized superfunctions would be
\begin{align}\nonumber\\[-28pt]
\raisebox{-29pt}{\includegraphics[scale=1]{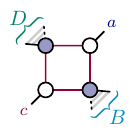}}&\equivR\raisebox{-24.65pt}{\includegraphics[scale=1]{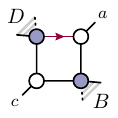}}{+}\raisebox{-24.65pt}{\includegraphics[scale=1]{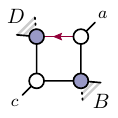}}\\[-5pt]
&\hspace{-50pt}=f_\sigma(\vec{p})\,\delta^{2\times\mathcal{N}}\!\big(C\!\cdot\!\tilde\eta\big)\!\Big[\hspace{-1pt}\delta^{2\times\bar{\mathcal{N}}}\!\big(D_{1}\!\!\cdot\!\theta\big){+}\delta^{2\times\bar{\mathcal{N}}}\!\big(D_{2}\!\!\cdot\!\theta\big)\hspace{-2pt}\Big]\nonumber\\[-24pt]\nonumber
\end{align}
where
\vspace{-5pt}\eq{\begin{split}
&\fwboxR{0pt}{\raisebox{-2.75pt}{$C\!\equivR\!\!\!\left(\rule{0pt}{14pt}\right.$}\!}\begin{array}{@{}c@{$\,$}|@{$\,$}c@{}c@{}c@{$\,$}|@{$$}c@{$$}|@{$\,$}c@{}c@{}c@{}}\multicolumn{8}{c}{}\\[-24pt]\multicolumn{1}{@{}c@{}}{\raisebox{9.25pt}{$\!\b{a}$}}&\multicolumn{3}{@{$\!$}c@{$\,$}}{\t{\overbrace{\phantom{\rule{41.5pt}{0pt}}}^{\text{{\normalsize$\t{B}$}}}}}&\multicolumn{1}{@{}c@{}}{\raisebox{9.25pt}{$\!\!\r{c}$}}&\multicolumn{3}{@{$\!$}c@{$\,$}}{\g{\overbrace{\phantom{\rule{41.5pt}{0pt}}}^{\text{{\normalsize$\g{D}$}}}}}\\[-3pt]
1&\t{\cdots}&\frac{\ab{\t{b}\,\r{c}}}{\ab{\b{a}\,\r{c}}}&\t{\cdots}&\dzero&\g{\cdots}&\frac{\ab{\g{d}\,\r{c}}}{\ab{\b{a}\,\r{c}}}&\g{\cdots}\\[-0.0pt]
\dzero&\t{\cdots}&\ab{\b{a}\,\t{b}}&\t{\cdots}&\ab{\b{a}\,\r{c}}&\g{\cdots}&\ab{\b{a}\,\g{d}}&\g{\cdots}
\end{array}\fwboxL{0pt}{\!\!\raisebox{-2.75pt}{$\left.\rule{0pt}{14pt}\right)$}}\;\\[-6pt]
&\fwboxR{0pt}{\raisebox{-2.75pt}{$D_1\!\!\equivR\!\!\!\left(\rule{0pt}{14pt}\right.$}\!}\begin{array}{@{}c@{$\,$}|@{$\,$}c@{}c@{}c@{$\,$}|@{$$}c@{$$}|@{$\,$}c@{}c@{}c@{}}\multicolumn{8}{c}{}\\[-24pt]\multicolumn{1}{@{}c@{}}{\raisebox{9.25pt}{$\phantom{\!\b{a}}$}}&\multicolumn{3}{@{$\!$}c@{$\,$}}{\phantom{\t{\overbrace{\phantom{\rule{41.5pt}{0pt}}}^{\text{{\normalsize$\t{B}$}}}}}}&\multicolumn{1}{@{}c@{}}{\phantom{\raisebox{9.25pt}{$\!\!\r{c}$}}}&\multicolumn{3}{@{$\!$}c@{$\,$}}{\phantom{\g{\overbrace{\phantom{\rule{41.5pt}{0pt}}}^{\text{{\normalsize$\g{D}$}}}}}}\\[-3pt]
\dzero&\dzero&{\color{dim}\cdots}&\dzero&\dzero&\g{\cdots}&\frac{\ab{\g{d}\,\r{c}}}{\ab{\b{a}\,\r{c}}}&\g{\cdots}\\[-0.0pt]
\dzero&\t{\cdots}&\ab{\b{a}\,\t{b}}&\t{\cdots}&\ab{\b{a}\,\r{c}}&\g{\cdots}&\ab{\b{a}\,\g{d}}&\g{\cdots}
\end{array}\fwboxL{0pt}{\!\!\raisebox{-2.75pt}{$\left.\rule{0pt}{14pt}\right)$}}\;\\[-6pt]
&
\fwboxR{0pt}{\raisebox{-2.75pt}{$D_2\!\!\equivR\!\!\!\left(\rule{0pt}{14pt}\right.$}\!}\begin{array}{@{}c@{$\,$}|@{$\,$}c@{}c@{}c@{$\,$}|@{$$}c@{$$}|@{$\,$}c@{}c@{}c@{}}\multicolumn{8}{c}{}\\[-24pt]\multicolumn{1}{@{}c@{}}{\raisebox{9.25pt}{$\phantom{\!\b{a}}$}}&\multicolumn{3}{@{$\!$}c@{$\,$}}{\phantom{\t{\overbrace{\phantom{\rule{41.5pt}{0pt}}}^{\text{{\normalsize$\t{B}$}}}}}}&\multicolumn{1}{@{}c@{}}{\phantom{\raisebox{9.25pt}{$\!\!\r{c}$}}}&\multicolumn{3}{@{$\!$}c@{$\,$}}{\phantom{\g{\overbrace{\phantom{\rule{41.5pt}{0pt}}}^{\text{{\normalsize$\g{D}$}}}}}}\\[-3pt]
1&\t{\cdots}&\frac{\ab{\t{b}\,\r{c}}}{\ab{\b{a}\,\r{c}}}&\t{\cdots}&\dzero&\dzero&{\color{dim}\cdots}&\dzero\\[-0.0pt]
\dzero&\t{\cdots}&\ab{\b{a}\,\t{b}}&\t{\cdots}&\ab{\b{a}\,\r{c}}&\g{\cdots}&\ab{\b{a}\,\g{d}}&\g{\cdots}
\end{array}\fwboxL{0pt}{\!\!\raisebox{-2.75pt}{$\left.\rule{0pt}{14pt}\right)\!.$}}\end{split}
\vspace{-8pt}}
Notice that $C\!\simeq\!\lambda$ as a 2-plane, so that $\mathrm{det}(c_a,c_b){=}\ab{a\,b}$ for all pairs $\{a,b\}\!\subset\![n]$; as such $\delta\big(C\cdot\tilde\eta\big){=}\hspace{1pt}\delta\big(\lambda\cdot\tilde\eta\big)$. Moreover, notice that $D_1$ and $D_2$ differ only in their top row; their bottom row is identical to that of $C$, and the sum of their top rows gives the top row of $C$. For each of the helicity configurations in (\ref{eg_mhv_one_loop_safe_components}), it is not hard to see that exactly one of the $\theta$-dependent $\delta$ functions contributes, giving exactly that which would arise from $\delta\big(\lambda\!\cdot\!\theta\big)$ (thereby ensuring equivalence to the corresponding component in the maximally supersymmetric case). And for the components (\ref{eg_mhv_one_loop_differing_components}) for which these functions differ, they result precisely in (\ref{general_n_2me_ls_olde}).

\vspace{-16pt}\subsection{Generalized Superfunctions at One Loop}\vspace{-14pt}
Let us now consider the general case of a box at one-loop, with $n$ external legs and N${}^{k{-}2}$MHV-degree:
\vspace{-12pt}\eq{\raisebox{-26pt}{\includegraphics[scale=1]{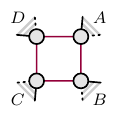}}\!\!\equivR\!\!\raisebox{-26pt}{\includegraphics[scale=1]{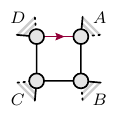}}\!\!{+}\!\!\raisebox{-26pt}{\includegraphics[scale=1]{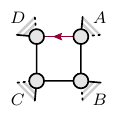}}\fwboxL{0pt}{.}\label{generalized_one_loop_pictures}
\vspace{-12pt}}
To find a generalized superfunction for this leading singularity, we first cut the marked edge to find
\vspace{-8pt}\eq{\raisebox{-12pt}{\includegraphics[scale=1]{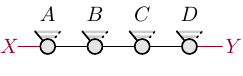}}\vspace{-8pt}}
which, being an tree of trees amplitudes, can always be expressed in terms of ordinary superfunctions of the form $f_\sigma(\vec{p})\,\delta^{(k\pl1)\times(4\sigma)}\!\big(\hat{C}\!\cdot\!\tilde\eta\big)$. As only the (ratios of) minors of the matrix $\hat{C}$ are relevant, it really should be understood as a representative of an equivalence class of $(k{+}1)\!\times\!(n{+}2)$ matrices---that is, as an element of $Gr(k{+}1,n{+}2)$. Let us choose a $GL(k{+}1)$-representative for $\hat{C}$ for which columns $c_{\r{X}}$ and $c_{\r{Y}}$ are taken as two of the basis elements:\\[-12pt]
\vspace{-10pt}\eq{\fwbox{0pt}{\raisebox{-25pt}{\hspace{-15pt}\includegraphics[scale=1]{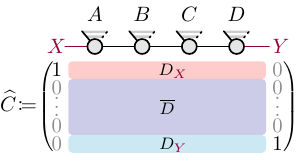}}.}\vspace{-6pt}\label{gauge_fixed_one_loop_pre_matrix}}
Summing over only those conjugate state pairs $|\r{X}\rangle,|\r{Y}\rangle$ relevant to an $\mathcal{N}$-supersymmetric theory according to (\ref{general_state_sum}), the $\theta$ integrations would result in minors either involving column $c_{\r{X}}$ or $c_{\r{Y}}$ \emph{exclusively}, and the $\tilde\eta$ integrations would result in minors involving either column \emph{inclusively}. Thus, upon summing over just the states that exist in an $\mathcal{N}$-supersymmetric theory, we obtain a superfunction of the form 
\vspace{-4pt}\eq{f_\sigma(\vec{p})\,\delta^{k\times\mathcal{N}}\!\big(C\!\cdot\!\tilde\eta\big)\!\Big[\hspace{-1pt}\delta^{k\times\bar{\mathcal{N}}}\!\big(D_{1}\!\!\cdot\!\theta\big){+}\delta^{k\times\bar{\mathcal{N}}}\!\big(D_{2}\!\!\cdot\!\theta\big)\hspace{-2pt}\Big]\vspace{-4pt}\label{general_one_loop_super}}
where 
\vspace{-4pt}\eq{C\!\equivR\!\!\!\left(\begin{array}{@{$\!$}c@{$\!$}}D_{\r{X}}{+}D_{\r{Y}}\\\bar{D}\end{array}\right)\!,\;D_1\!\equivR\!\!\!\left(\begin{array}{@{$\,$}l@{$\!$}}\bar{D}\\\hspace{-7.5pt}\phantom{{}_{{Y}}}D_{\r{{Y}}}\hspace{-10pt}\end{array}\right)\!,\;D_2\!\equivR\!\!\!\left(\begin{array}{@{$\,$}l@{$\!$}}\hspace{-7.5pt}\phantom{{}_{{X}}}D_{\r{X}}\hspace{-10pt}\\\bar{D}\end{array}\right)\!.\vspace{-4pt}}
More invariantly, Pl\"ucker coordinates for these particular elements of $Gr(k,n)$ can be defined by
\begin{align}\nonumber\\[-20pt]
\fwboxR{60pt}{\Delta_{C_{\phantom{1\!}}}[a_1,\ldots,a_k]}&\!\equivR\!\!\Delta_{\hat{C}}[\r{X},a_1,\ldots,a_k]{+}\Delta_{\hat{C}}[a_1,\ldots,a_k,\r{Y}]\nonumber\\
\fwboxR{60pt}{\Delta_{D_1\!}[a_1,\ldots,a_k]}&\!\equivR\!\!\Delta_{\hat{C}}[\r{X},a_1,\ldots,a_k]\label{pluckers_for_projections}\\
\fwboxR{60pt}{\Delta_{D_2\!}[a_1,\ldots,a_k]}&\!\equivR\!\!\Delta_{\hat{C}}[a_1,\ldots,a_k,\r{Y}].\nonumber\\[-20pt]\nonumber
\end{align}
The first of these rules defines the `merge' operation in maximal supersymmetry \cite{ArkaniHamed:book}. Notice that as all three matrices $C,D_1,D_2$ overlap on the $(k{-}1)$-dimensional subspace $\bar{D}$, we can alternatively write (\ref{general_one_loop_super}) as
\vspace{-2pt}\eq{\begin{split}f_\sigma(\vec{p})\,\delta^{(k\mi1)\times(4\sigma)}\!\big(\bar{D}\!\cdot\!\tilde\eta\big)&\delta^{\mathcal{N}}\!\big((D_{\r{X}}{+}D_{\r{Y}})\!\cdot\!\tilde\eta\big)\\
&\fwboxR{0pt}{\times}\Big[\hspace{-0pt}\delta^{\bar{\mathcal{N}}}\!\big(D_{\r{X}}\!\cdot\!\theta\big){+}\delta^{\bar{\mathcal{N}}}\!\big(D_{\r{Y}}\!\cdot\!\theta\big)\hspace{-0pt}\Big]\end{split}\vspace{-15pt}}
which shows that, relative to the maximally supersymmetric case, generalized superfunctions at one loop differ by a single row of their Grassmann $\delta$-functions. 
\vspace{-16pt}\subsection{\emph{Exempli Gratia}: NMHV Leading Singularities}\vspace{-14pt}
It may be worthwhile to consider one more example at one loop in detail. Consider the case of a three-mass box leading singularity relevant to NMHV amplitudes:
\begin{align}\nonumber\\[-28pt]
\raisebox{-29pt}{\includegraphics[scale=1]{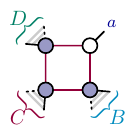}}&\equivR\raisebox{-24.65pt}{\includegraphics[scale=1]{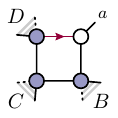}}{+}\raisebox{-24.65pt}{\includegraphics[scale=1]{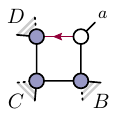}}\\[-5pt]
&\hspace{-50pt}=f_\sigma(\vec{p})\,\delta^{3\times\mathcal{N}}\!\big(C\!\cdot\!\tilde\eta\big)\!\Big[\hspace{-1pt}\delta^{3\times\bar{\mathcal{N}}}\!\big(D_{1}\!\!\cdot\!\theta\big){+}\delta^{3\times\bar{\mathcal{N}}}\!\big(D_{2}\!\!\cdot\!\theta\big)\hspace{-2pt}\Big];\nonumber\\[-24pt]\nonumber
\end{align}
Here, the ordinary function $f_1(\vec{p})$ in the case of (plane-ordered) Yang-Mills\footnote{As before, the corresponding function $f_{2}(\vec{p})$ for gravity is not hard to construct from the relevant tree amplitudes.} is given by 
\vspace{-5pt}\eq{f_1(\vec{p})\!\equivR\!\!\frac{\hspace{-0pt}\ab{\hspace{-1pt}\t{b_{\text{-}1}}\r{c_{1}}\hspace{-1pt}}\!\ab{\hspace{-1pt}\r{c_{\text{-}1}}\g{d_1}\hspace{-1pt}}\fwboxL{0pt}{\delta^{2\times2}\!\big(\lambda\!\cdot\!\tilde\lambda\big)
}}{\ab{12}\!\cdots\!\ab{n1}\!\ab{\hspace{-1pt}\b{a}|\hspace{-1pt}\t{B}\hspace{-1pt}|\hspace{-1pt}\r{C}\hspace{-1pt}|\hspace{-1pt}\g{d_{1}}\hspace{-1pt}}\!\ab{\hspace{-1pt}\b{a}\hspace{-1pt}|\hspace{-1pt}\t{B}\hspace{-1pt}|\hspace{-1pt}\r{C}\hspace{-1pt}|\hspace{-1pt}\r{c_{\text{-}1}}\hspace{-1pt}}\!\ab{\hspace{-1pt}\t{b_{\text{-}1}}\hspace{-1pt}|\hspace{-1pt}\r{C}\hspace{-1pt}|\hspace{-1pt}\g{D}\hspace{-1pt}|\hspace{-1pt}\b{a}\hspace{-1pt}}\!\ab{\hspace{-1pt}\r{c_1}\hspace{-1pt}|\hspace{-1pt}\r{C}\hspace{-1pt}|\hspace{-1pt}\g{D}\hspace{-1pt}|\hspace{-1pt}\b{a}\hspace{-1pt}}\hspace{-1pt}\r{s_C}}\nonumber\vspace{-4pt}}
and the matrices may be written
\vspace{5pt}\eq{\begin{split}
&\hspace{-39pt}\fwboxR{0pt}{\raisebox{-2.75pt}{$C\!\equivR\!\!\!\left(\rule{0pt}{22pt}\right.$}\!}\begin{array}{@{}c@{$\,$}|@{$\,$}c@{}c@{}c@{$\,$}|@{$\,$}c@{}c@{}c@{$\,$}|@{$\,$}c@{}c@{}c@{}}\multicolumn{10}{c}{}\\[-24pt]\multicolumn{1}{@{}c@{}}{\raisebox{9.25pt}{$\!\!\b{a}$}}&\multicolumn{3}{@{$\!$}c@{$\,$}}{\t{\overbrace{\phantom{\rule{61.5pt}{0pt}}}^{\text{{\normalsize$\t{B}$}}}}}&\multicolumn{3}{@{$\!$}c@{$\,$}}{\r{\overbrace{\phantom{\rule{61.5pt}{0pt}}}^{\text{{\normalsize$\r{C}$}}}}}&\multicolumn{3}{@{$\!$}c@{$\,$}}{\g{\overbrace{\phantom{\rule{57.5pt}{0pt}}}^{\text{{\normalsize$\g{D}$}}}}}\\[-3pt]
1&\b{\cdots}&\frac{\ab{\hspace{-1pt}\t{b}|\r{C}|\g{D}|\b{a}\hspace{-1pt}}}{\ab{\hspace{-1pt}\b{a}|\r{C}|\g{D}|\hspace{-1pt}\b{a}}}&\b{\cdots}&\r{\cdots}&\frac{\ab{\hspace{-1pt}\r{c}|\r{C}|\g{D}|\b{a}\hspace{-1pt}}}{\ab{\hspace{-1pt}\b{a}|\r{C}|\g{D}|\hspace{-1pt}\b{a}}}&\r{\cdots}&\g{\cdots}&\frac{\ab{\hspace{-1pt}\g{d}|\r{C}|\g{D}|\b{a}\hspace{-1pt}}}{\ab{\hspace{-1pt}\b{a}|\r{C}|\g{D}|\hspace{-1pt}\b{a}}}&\g{\cdots}\\[-1pt]
\dzero&\t{\cdots}&\ab{\hspace{-0pt}\b{a}\,\t{b}\hspace{-0pt}}&\t{\cdots}&\r{\cdots}&\ab{\hspace{-0pt}\b{a}\,\r{c}\hspace{-0pt}}&\r{\cdots}&\g{\cdots}&\ab{\hspace{-0pt}\b{a}\,\g{d}\hspace{-0pt}}&\g{\cdots}\\[0.75pt]
\dzero&\t{\cdots}&\ab{\hspace{-1pt}\b{a}\,\t{b}\hspace{-1pt}}\hspace{-1pt}\r{s_{\r{C}}}&\t{\cdots}&\r{\cdots}&\ab{\hspace{-1pt}\b{a}\hspace{-1pt}|\hspace{-1pt}\t{B}\hspace{-1pt}|\hspace{-1pt}\r{C}\hspace{-1pt}|\hspace{-1pt}\r{c}\hspace{-1pt}}&\r{\cdots}&\dzero&{\color{dim}\cdots}&\dzero
\end{array}\fwboxL{0pt}{\!\!\raisebox{-2.75pt}{$\left.\rule{0pt}{22pt}\right)$}}\hspace{-50pt}\\[-2pt]
&\hspace{-39pt}\fwboxR{0pt}{\raisebox{-2.75pt}{$D_1\!\!\equivR\!\!\!\left(\rule{0pt}{22pt}\right.$}\!}\begin{array}{@{}c@{$\,$}|@{$\,$}c@{}c@{}c@{$\,$}|@{$\,$}c@{}c@{}c@{$\,$}|@{$\,$}c@{}c@{}c@{}}
\multicolumn{10}{c}{}\\[-24pt]\multicolumn{1}{@{}c@{}}{\phantom{\raisebox{9.25pt}{$\b{a}$}}}&\multicolumn{3}{@{$\!$}c@{$\,$}}{\phantom{\t{\overbrace{\phantom{\rule{62.5pt}{0pt}}}^{\text{{\normalsize$\t{B}$}}}}}}&\multicolumn{3}{@{$\!$}c@{$\,$}}{\phantom{\r{\overbrace{\phantom{\rule{62.5pt}{0pt}}}^{\text{{\normalsize$\r{C}$}}}}}}&\multicolumn{3}{@{$\!$}c@{$\,$}}{\phantom{\g{\overbrace{\phantom{\rule{57.5pt}{0pt}}}^{\text{{\normalsize$\g{D}$}}}}}}\\[-3pt]
\dzero&\dzero&{\color{dim}\cdots}&\dzero&\r{\cdots}&\frac{\ab{\hspace{-1pt}\r{c}|\r{C}|\g{D}|\b{a}\hspace{-1pt}}}{\ab{\hspace{-1pt}\b{a}|\r{C}|\g{D}|\hspace{-1pt}\b{a}}}&\r{\cdots}&\g{\cdots}&\frac{\ab{\hspace{-1pt}\g{d}|\r{C}|\g{D}|\b{a}\hspace{-1pt}}}{\ab{\hspace{-1pt}\b{a}|\r{C}|\g{D}|\hspace{-1pt}\b{a}}}&\g{\cdots}\\[-1pt]
\dzero&\t{\cdots}&\ab{\hspace{-0pt}\b{a}\,\t{b}\hspace{-0pt}}&\t{\cdots}&\r{\cdots}&\ab{\hspace{-0pt}\b{a}\,\r{c}\hspace{-0pt}}&\r{\cdots}&\g{\cdots}&\ab{\hspace{-0pt}\b{a}\,\g{d}\hspace{-0pt}}&\g{\cdots}\\[0.75pt]
\dzero&\t{\cdots}&\ab{\hspace{-1pt}\b{a}\,\t{b}\hspace{-1pt}}\hspace{-1pt}\r{s_{\r{C}}}&\t{\cdots}&\r{\cdots}&\ab{\hspace{-1pt}\b{a}\hspace{-1pt}|\hspace{-1pt}\t{B}\hspace{-1pt}|\hspace{-1pt}\r{C}\hspace{-1pt}|\hspace{-1pt}\r{c}\hspace{-1pt}}&\r{\cdots}&\dzero&{\color{dim}\cdots}&\dzero
\end{array}\fwboxL{0pt}{\!\!\raisebox{-2.75pt}{$\left.\rule{0pt}{22pt}\right)$}}\hspace{-50pt}\\[-6pt]
&\hspace{-39pt}\fwboxR{0pt}{\raisebox{-2.75pt}{$D_2\!\!\equivR\!\!\!\left(\rule{0pt}{22pt}\right.$}\!}\begin{array}{@{}c@{$\,$}|@{$\,$}c@{}c@{}c@{$\,$}|@{$\,$}c@{}c@{}c@{$\,$}|@{$\,$}c@{}c@{}c@{}}
\multicolumn{10}{c}{}\\[-24pt]\multicolumn{1}{@{}c@{}}{\phantom{\raisebox{9.25pt}{$\b{a}$}}}&\multicolumn{3}{@{$\!$}c@{$\,$}}{\phantom{\t{\overbrace{\phantom{\rule{62.5pt}{0pt}}}^{\text{{\normalsize$\t{B}$}}}}}}&\multicolumn{3}{@{$\!$}c@{$\,$}}{\phantom{\r{\overbrace{\phantom{\rule{62.5pt}{0pt}}}^{\text{{\normalsize$\r{C}$}}}}}}&\multicolumn{3}{@{$\!$}c@{$\,$}}{\phantom{\g{\overbrace{\phantom{\rule{57.5pt}{0pt}}}^{\text{{\normalsize$\g{D}$}}}}}}\\[-3pt]
1&\t{\cdots}&\frac{\ab{\hspace{-1pt}\t{b}|\r{C}|\g{D}|\b{a}\hspace{-1pt}}}{\ab{\hspace{-1pt}\b{a}|\r{C}|\g{D}|\hspace{-1pt}\b{a}}}&\t{\cdots}&\dzero&{\color{dim}\cdots}&\dzero&\dzero&{\color{dim}\cdots}&\dzero\\[-1pt]
\dzero&\t{\cdots}&\ab{\hspace{-0pt}\b{a}\,\t{b}\hspace{-0pt}}&\t{\cdots}&\r{\cdots}&\ab{\hspace{-0pt}\b{a}\,\r{c}\hspace{-0pt}}&\r{\cdots}&\g{\cdots}&\ab{\hspace{-0pt}\b{a}\,\g{d}\hspace{-0pt}}&\g{\cdots}\\[0.75pt]
\dzero&\t{\cdots}&\ab{\hspace{-1pt}\b{a}\,\t{b}\hspace{-1pt}}\hspace{-1pt}\r{s_{\r{C}}}&\t{\cdots}&\r{\cdots}&\ab{\hspace{-1pt}\b{a}\hspace{-1pt}|\hspace{-1pt}\t{B}\hspace{-1pt}|\hspace{-1pt}\r{C}\hspace{-1pt}|\hspace{-1pt}\r{c}\hspace{-1pt}}&\r{\cdots}&\dzero&{\color{dim}\cdots}&\dzero
\end{array}\fwboxL{0pt}{\!\!\raisebox{-2.75pt}{$\left.\rule{0pt}{22pt}\right)\!.$}}\hspace{-50pt}
\end{split}
}
As before, we notice that all three matrices share their bottom two rows and the first rows of $D_1$ and $D_2$ combine to that of $C$; finally, notice that the top two rows of $C$ are again equivalent to the two-plane $\lambda$ (as required by super-momentum conservation).

\vspace{-16pt}\section{Generalized Superfunctions Beyond One Loop}\vspace{-14pt}
This story naturally generalizes to arbitrary loop order, resulting in the general form given in (\ref{archetype_less_superfunction}). Here, the matrix $C$ is always what would have been found for the leading singularity in the maximally supersymmetric theory, and the $D_{\vec{s}}$ are defined by some choice of projecting out exactly one of $\{\r{X_i},\r{Y_i}\}$ for each $i\!\in\![L]$. The only interesting distinction relative to one loop is that, beyond some certain loop order depending on N${}^{k\mi2}$MHV-degree, it is not generally possible to choose a representative $\hat{C}$-matrix like in (\ref{gauge_fixed_one_loop_pre_matrix}), as $2L$ may exceed $k{+}L$; when this happens, it will no longer be the case that all $D_{\vec{s}}$ are identical along a subsets of rows.

More interestingly, perhaps, beyond one loop is the appearance of \emph{non-polylogarithmic} leading singularities starting at two loops (see~\emph{e.g.}~\cite{Bourjaily:2017bsb,Bourjaily:2018yfy,Bourjaily:2018ycu,Bourjaily:2021vyj,Bourjaily:2022tep,Bourjaily:2022bwx}). These are not defined directly in terms of the products of amplitudes localized by on-shell conditions and momentum conservation, but over particular algebraic varieties defined within the on-shell phase-space of these internal particles. Nevertheless, even these leading singularities can be cut into superfunctions defined by trees of tree-amplitudes; the real concern is that the contours over which they are defined may depend on components in general, resulting in potentially large differences among different contributions. We leave the investigation of such leading singularities to future work.

\vspace{-16pt}\section{Conclusions and Future Directions}\vspace{-14pt}
In terms of generalized superfunctions, it is much easier to obtain closed analytic expressions for one loop amplitudes in sYM and sGR for arbitrary multiplicity and N$^k$MHV-degree. We hope that these will soon give way to analytic expressions for all one-loop amplitudes in pure (or sub-maximally supersymmetric) gauge theory and gravity, greatly expanding what is known today.\\ 

While this formalism should suffice for the entire so-called cut-constructible part of loop amplitudes in pure gauge theory or gravity, it would be valuable to have a better understanding of what is required by regularized Feynman rules to determine complete (regulated) amplitudes. While there is some evidence that strictly four-dimensional data can faithfully capture at least some of this information (for example, the all-plus amplitudes at one loop \cite{Henn:2019mvc}), it remains to be seen why this works, when it does, or how such ideas may be generalized. Luckily, there is a growing body of data on the form taken by these contributions to amplitudes in pure gauge theory (see, \emph{e.g.}~\mbox{\cite{Bern:2005ji,Elvang:2011ub,Badger:2015lda,Dunbar:2016gjb,Badger:2019djh}}).

\newpage
\section*{Acknowledgements}\vspace{-14pt}
The author acknowledges fruitful conversations with Song He, Johannes Henn, Michael Plesser, Radu Roiban, Jaroslav Trnka, and Philip Velie. This work was supported in part by a grant from the US Department of Energy (No.\ DE-SC00019066).

\vspace{-14pt}
\providecommand{\href}[2]{#2}\begingroup\raggedright\endgroup

\end{document}